\documentclass[aps,prc,twocolumn,superscriptaddress,showpacs,amsmath,floatfix,
nofootinbib]{revtex4}
\usepackage{graphicx}
\usepackage{bm}

\begin{document}


\title{Shell model study of single-particle and collective structure
in neutron-rich Cr isotopes}

\author{K.~Kaneko}
\email{kaneko@ip.kyusan-u.ac.jp} \affiliation{Department of Physics,
Kyushu Sangyo University, Fukuoka 813-8503, Japan}
\author{Y. Sun}
\email{sunyang@sjtu.edu.cn} \affiliation{Department of Physics,
Shanghai Jiao Tong University, Shanghai 200240, P. R. China}
\author{M. Hasegawa}
\affiliation{Institute of Modern Physics, Chinese Academy of
Sciences, Lanzhou 730000, P. R. China}
\author{T. Mizusaki}
\affiliation{Institute of Natural Sciences, Senshu University, Tokyo
101-8425, Japan}

\begin{abstract}

The structure of neutron-rich Cr isotopes is systematically
investigated by using the spherical shell model. The calculations
reproduce well the known energy levels for the even-even
$^{52-62}$Cr and odd-mass $^{53-59}$Cr nuclei, and predict a
lowering of excitation energies around neutron number $N=40$. The
calculated $B(E2;2_{1}^{+}\rightarrow 0_{1}^{+})$ systematics
shows a pronounced collectivity around $N=40$; a similar
characteristic behavior has been suggested for Zn and Ge isotopes.
Causes for the sudden drop of the $9/2_{1}^{+}$ energy in
$^{59}$Cr and the appearance of very low $0_{2}^{+}$ states around
$N=40$ are discussed. We also predict a new band with strong
collectivity built on the $0_{2}^{+}$ state in the $N=40$ isotope
$^{64}$Cr.

\end{abstract}


\maketitle

\section{Introduction}\label{sec1}

The neutron-rich $fp$-shell nuclei far from the valley of
stability are of particular interest in recent experimental and
theoretical studies \cite{Janssens05}. To provide a satisfactory
description of these nuclei, the challenge remains to understand
what mechanisms cause changes in nuclear shell structure as
neutron number increases in nuclear systems. Theoretical
calculations have questioned the persistence of the traditional
magic numbers, which have been known to exist in stable nuclei.
For example, $^{68}$Ni is expected to be a double-magic nucleus as
a consequence of the neutron subshell gap separating the
$fp$-shell and $g_{9/2}$ orbital at the neutron number $N=40$
\cite{Broda,Grzywacz98,Ishii}. The size of this subshell gap is
characterized by the excitation across the $fp$-shell and
$g_{9/2}$ orbital \cite{Sorlin02,Kaneko06}. In addition to this,
the presence of the positive-parity $g_{9/2}$ orbital above the
negative-parity $fp$ shell strongly hinders 1p-1h excitations.
This $N=40$ shell gap, however, disappears when protons are added
to or subtracted from $^{68}$Ni \cite{Hannawald}. It was discussed
in Refs. \cite{Reinhard,Grawe} that in the early mean-field
calculations, a distinct shell gap that exists in the $N=$ 40
nucleus $^{68}$Ni disappears when quadrupole correlations are
taken into account. Even for $^{68}$Ni, it does not show a
pronounced irregularity in the two-neutron separation energy as
expected for a typical double-magic nucleus. It has been suggested
\cite{Langanke} that a small $B(E2,0_{1}^{+}\rightarrow
2_{1}^{+})$ value is not a strong evidence for the double-magic
character. We may thus conclude that the double-magicity nature in
$^{68}$Ni is still controversial and remains an open question.

On the other hand, collective structure is predicted to develop in
neutron-rich nuclei. Zn and Ge nuclei are known to exhibit rather
strong collectivity around $N=40$ \cite{Padilla,Walle}. Both
unusually low excitation of the first excited $0_{2}^{+}$ state
and strong enhancement of $B(E2;0_{1}^{+}\rightarrow 2_{1}^{+})$
near $N=40$ indicate a dramatic structure change at this nucleon
number. Recently, we have shown
\cite{Hasegawa07} that this characteristic behavior can possibly
be understood in terms of rapid increase in the $g_{9/2}$ proton
and neutron occupation. However, in order to explain the structure
change and enhancement in $B(E2)$, we needed effective quadrupole
matrix elements and large effective charges. Shell model
calculations \cite{Kaneko06} have also predicted an excited band
in $^{68}$Ni based on the $0_{2}^{+}$ state.

For $^{60,62,64}$Cr, shell-model calculations in the $fpg$-shell
space with the $^{48}$Ca core (i.e. the $^{40}$Ca core and eight
$f_{7/2}$ frozen neutrons) have shown \cite{Sorlin03} that these
nuclei are strongly deformed with large quadrupole moments and large
$B(E2;2_{1}^{+}\rightarrow 0_{1}^{+})$ values. The recent report
\cite{Zhu} on some even-even Cr isotopes has indicated a lowering of
the $2^{+}$ energy beyond $N=34$. The monopole interaction discussed
by Otsuka {\it et al.} \cite{Otsuka01,Otsuka05} should affect
differently the protons occupying the $\pi f_{7/2}$ and $\nu
f_{5/2}$ orbitals, and should have an impact on the neutron
single-particle spectra involved. However, it has been demonstrated
that the $\nu g_{9/2}$ orbital should not be ignored in the
description of $N\ge 34$ isotopes. In fact, the GXPF1A interaction
\cite{Honma} in the $fp$-shell space cannot describe the energy
levels in $^{59}$Cr and $^{60}$Cr \cite{Zhu}, and the calculated
excitation energies are found much higher than the experimental
ones. Sorlin {\it et al.} \cite{Sorlin02} suggested an onset of a
sizable deformation in $^{60}$Cr. Recent observation \cite{Zhu},
however, has questioned this interpretation in terms of deformation.
Rather, the low-lying energy systematics of neutron-rich Cr isotopes
has been discussed in terms of a softening of nuclear shape with
increasing neutron number.

The low-lying $9/2_{1}^{+}$ states in odd-mass neutron-rich Cr
isotopes are obviously outside the $fp$-shell model space. The
recent observation \cite{Freeman} in some odd-mass neutron-rich Cr
isotopes reveals that the $9/2_{1}^{+}$ state energy drops down
considerably with increasing neutron number. The sharp decrease of
the $9/2_{1}^{+}$ state energy is considered as a clear indication
for the monopole interaction, and the attractive force between the
$1g_{9/2}$ and $2p_{3/2}$ neutron orbitals pushes down the
$1g_{9/2}$ orbital. The mean-field calculations suggested the
presence of prolately-deformed rotational bands built on the
$g_{9/2}$ states in $^{55,57}$Cr, while the long-lived $g_{9/2}$
isomeric state observed in $^{59}$Cr was considered to be
consistent with an oblate deformation \cite{Deacon,Grzywacz99}.
All these indicate that different shapes may coexist and a shape
change may occur in $^{59}$Cr. The abrupt change in structure of
odd-mass neutron-rich Cr isotopes can be better explained by the
appearance of the $g_{9/2}$ orbital at low energy. The observation
of the $9/2_{1}^{+}$ state at low-excitation energy undoubtedly
demonstrates the necessity of including the $g_{9/2}$ orbital in
the calculation.

When these neutron-rich nuclei rotate faster, it is required also
to involve the $g_{9/2}$ orbital to discuss the physics. The study
of high-spin states in the even-even $^{56,58,60}$Cr isotopes is
important to gain information about the evolution of
single-particle and collective excitations. A comparison of the
high-spin data with current shell model calculations indicates
that there is strong motivation for extending the interactions to
include the neutron $g_{9/2}$ orbital \cite{Zhu}. Thus, a
consistent approach to the structure study in this mass region
would suggest that the $g_{9/2}$ orbital should drive toward
collectivity in the neutron-rich Cr isotopes for $N\ge 34$.

\begin{figure}[t]
\includegraphics[totalheight=9.5cm]{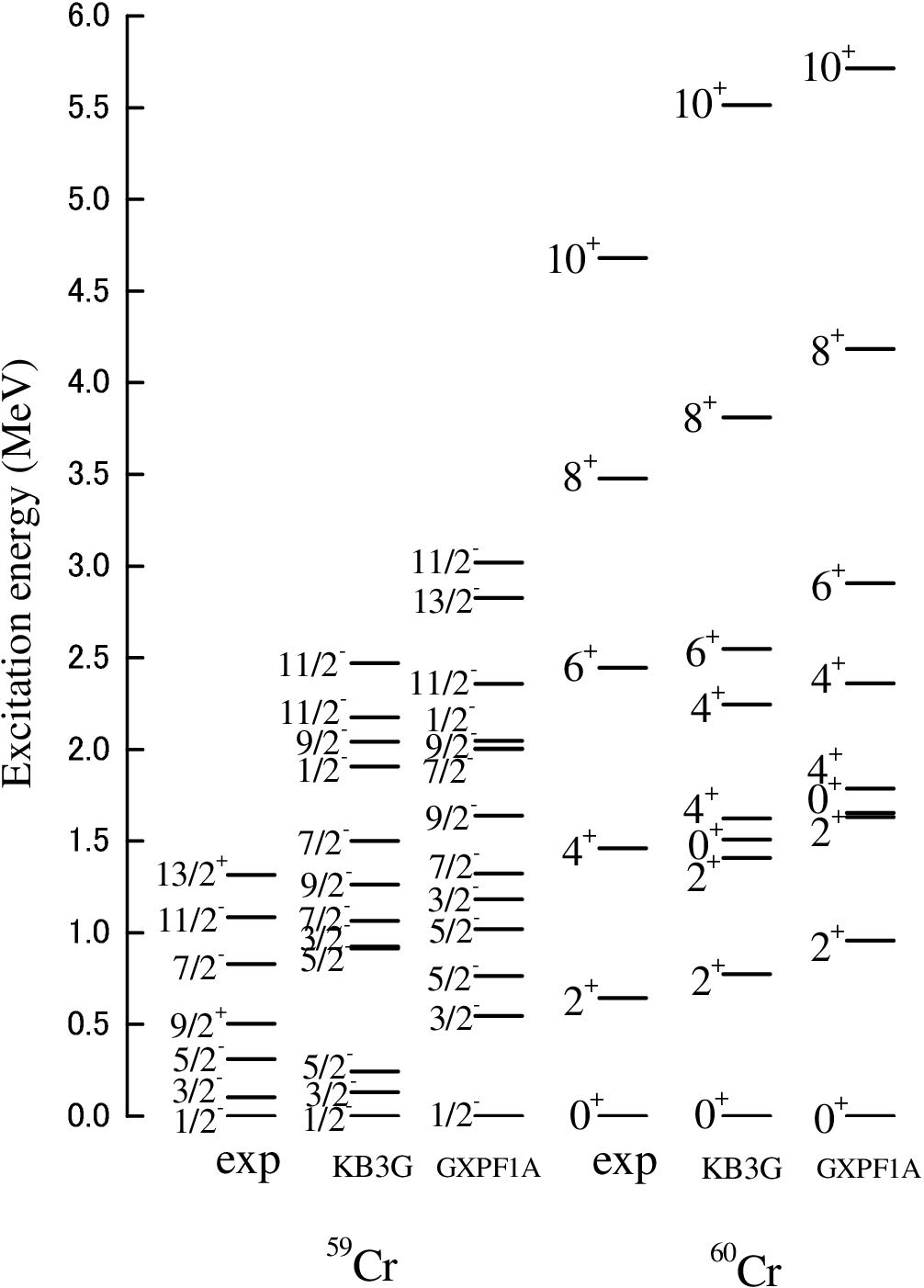}
\caption{Comparison of experimental energy levels
with the KB3G and the GXPF1A calculations for $^{59,60}$Cr.}
 \label{fig5a}
\end{figure}

In Fig. \ref{fig5a}, we compare the experimental energy levels of
$^{59}$Cr and $^{60}$Cr with the present calculation, and with
calculations using the GXPF1A and KB3G interactions. It has recently
been discussed \cite{Zhu} that the GXPF1A interaction \cite{Honma}
is inadequate to describe the neutron-rich Cr isotopes beyond
$N=34$. For example, for $^{59}$Cr their calculated energy levels of
the low-lying negative-parity states lie much higher than what are
seen in the data, and the discrepancy is further amplified in higher
spin yrast states. While their results for the $N\le 34$ Cr isotopes
reproduce the data, the agreement with the even-even $^{60}$Cr data
is poor for high-spin states. To understand the single-particle and
collective structure in neutron-rich Cr isotopes, shell-model
calculations in a full $fpg$-shell space are highly desirable.
However, conventional shell-model calculations in the full
$fpg$-shell space are not possible at present because of too huge
dimensions in the configuration space. For neutron-rich Cr isotopes,
we have to restrict the model space to the $fpg$ model space of
$^{48}$Ca, namely, using the $^{40}$Ca core with eight $f_{7/2}$
frozen neutrons. In the $N=40$ region, the $d_{5/2}$ orbital may
need to be taken into account to develop quadrupole collectivity
\cite{Hasegawa07,Sorlin03,Zuker,Caurier}. However, recent
observations in $^{80}$Zn \cite{Walle} and $^{82}$Ge \cite{Padilla}
indicate the persistence of the $N=50$ shell gap, implying that the
$d_{5/2}$ orbital may not affect much the collectivity of the
neutron-rich Cr isotopes with $N\le 50$.

In the present work, we perform large-scale spherical shell model
calculations using the pairing plus multipole forces with the
monopole interaction included \cite{Hasegawa01,Kaneko02}. The $fpg$
model space comprises the $0f_{7/2}$, $1p_{3/2}$, $0f_{5/2}$,
$1p_{1/2}$ active proton orbitals and $0f_{7/2}$, $1p_{3/2}$,
$0f_{5/2}$, $1p_{1/2}$, $0g_{9/2}$ neutron orbitals with eight
$f_{7/2}$ frozen neutrons. Structure of neutron-rich Cr isotopes
will be investigated in detail. In particular, the interplay between
single-particle and collective properties will be discussed.

The paper is arranged as follows. In Section~\ref{sec2}, we outline
our model. In Section~\ref{sec3}, we present the results from
numerical calculations and discuss them for a long chain of Cr
isotopes. Finally, conclusions are drawn in Section~\ref{sec5}.

\begin{figure}[t]
\includegraphics[totalheight=9.5cm]{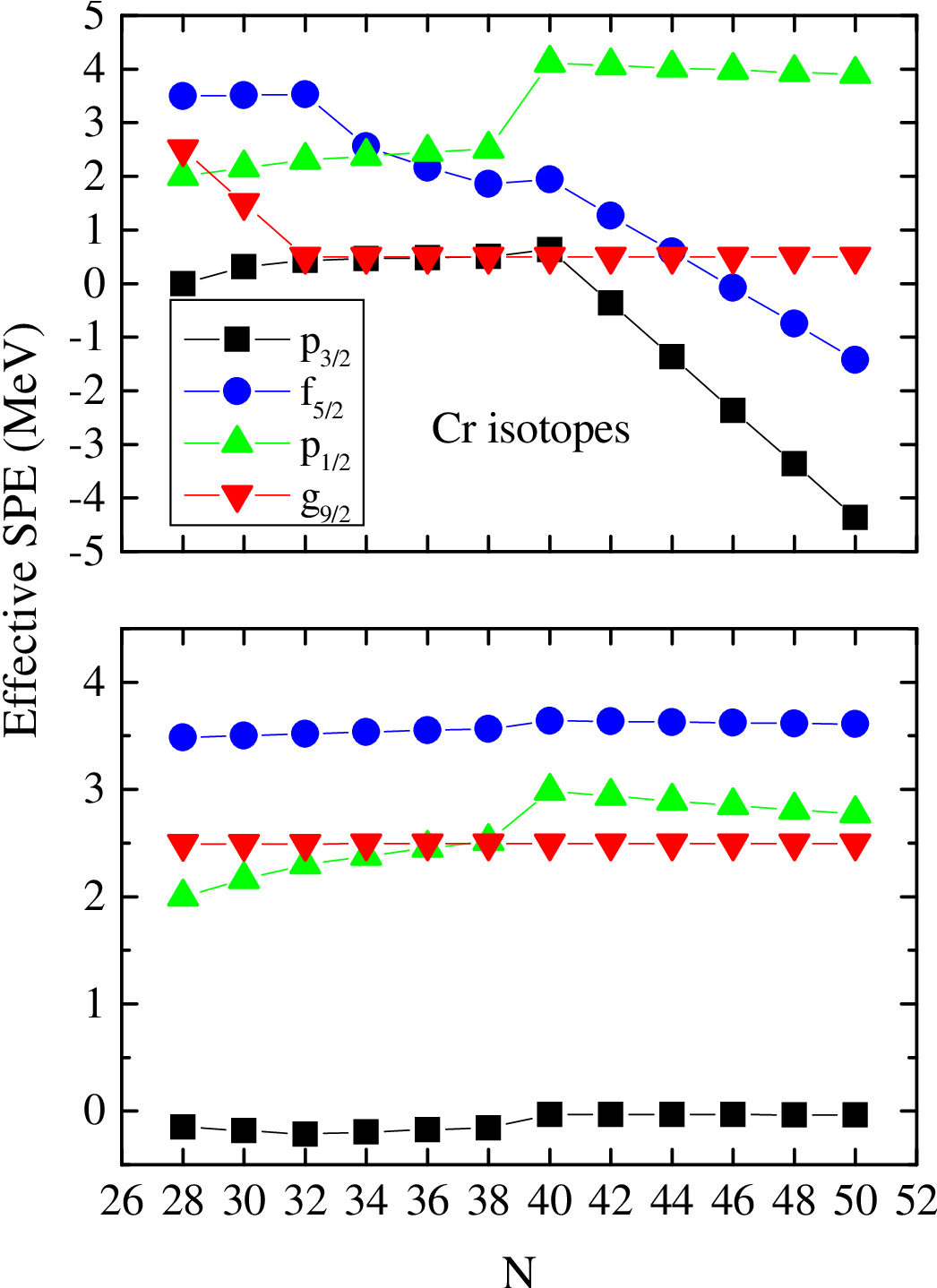}
\caption{(color online) Effective single-particle energies
$e_{\alpha}$ for even-even Cr isotopes. (Upper panel) calculation
from the shell-model Hamiltonian with the monopole interactions, and
(lower panel) without the monopole interactions.}
  \label{fig11}
\end{figure}
\begin{figure*}[htb]
\includegraphics[width=0.47\textwidth]{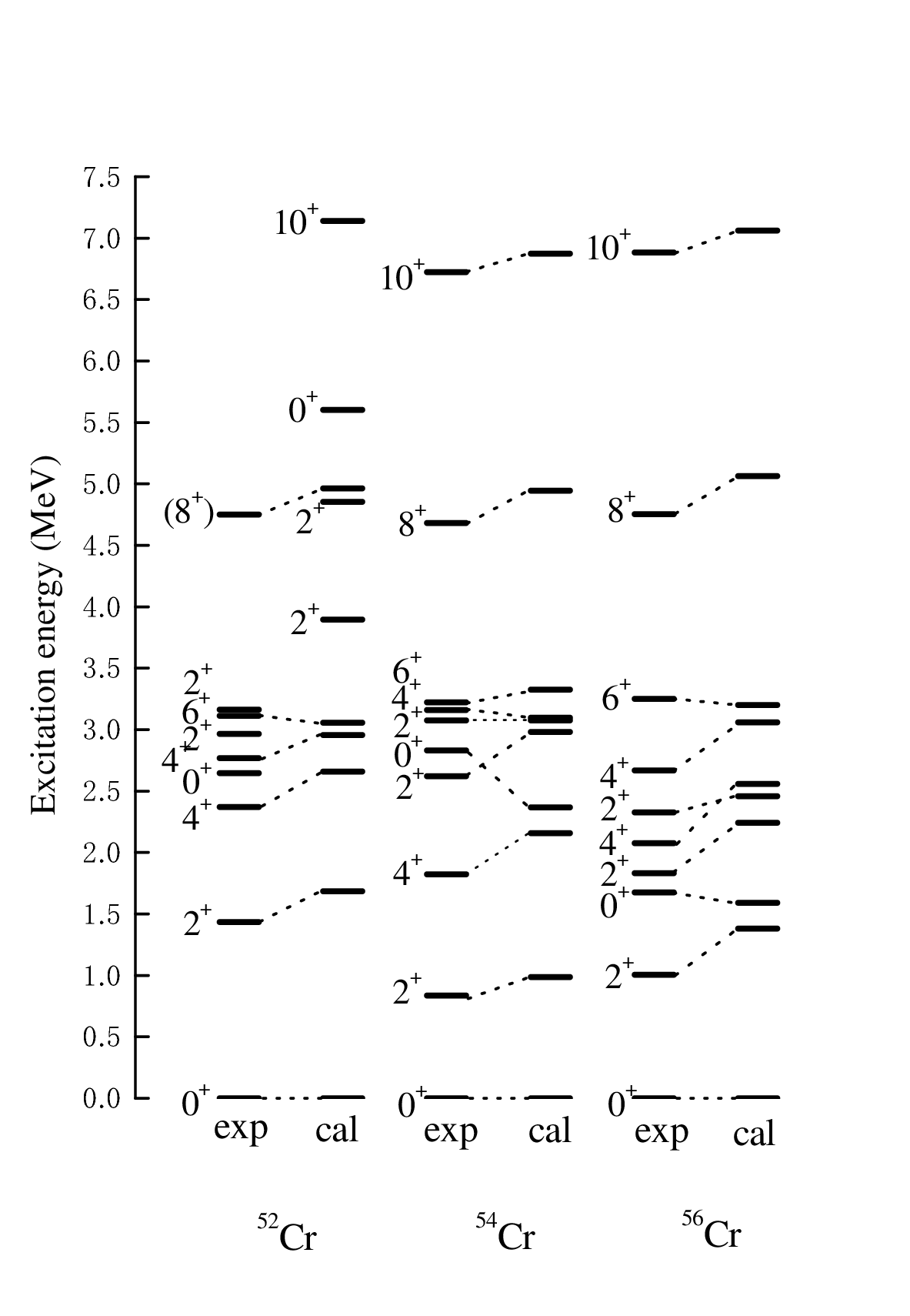}
\includegraphics[width=0.47\textwidth]{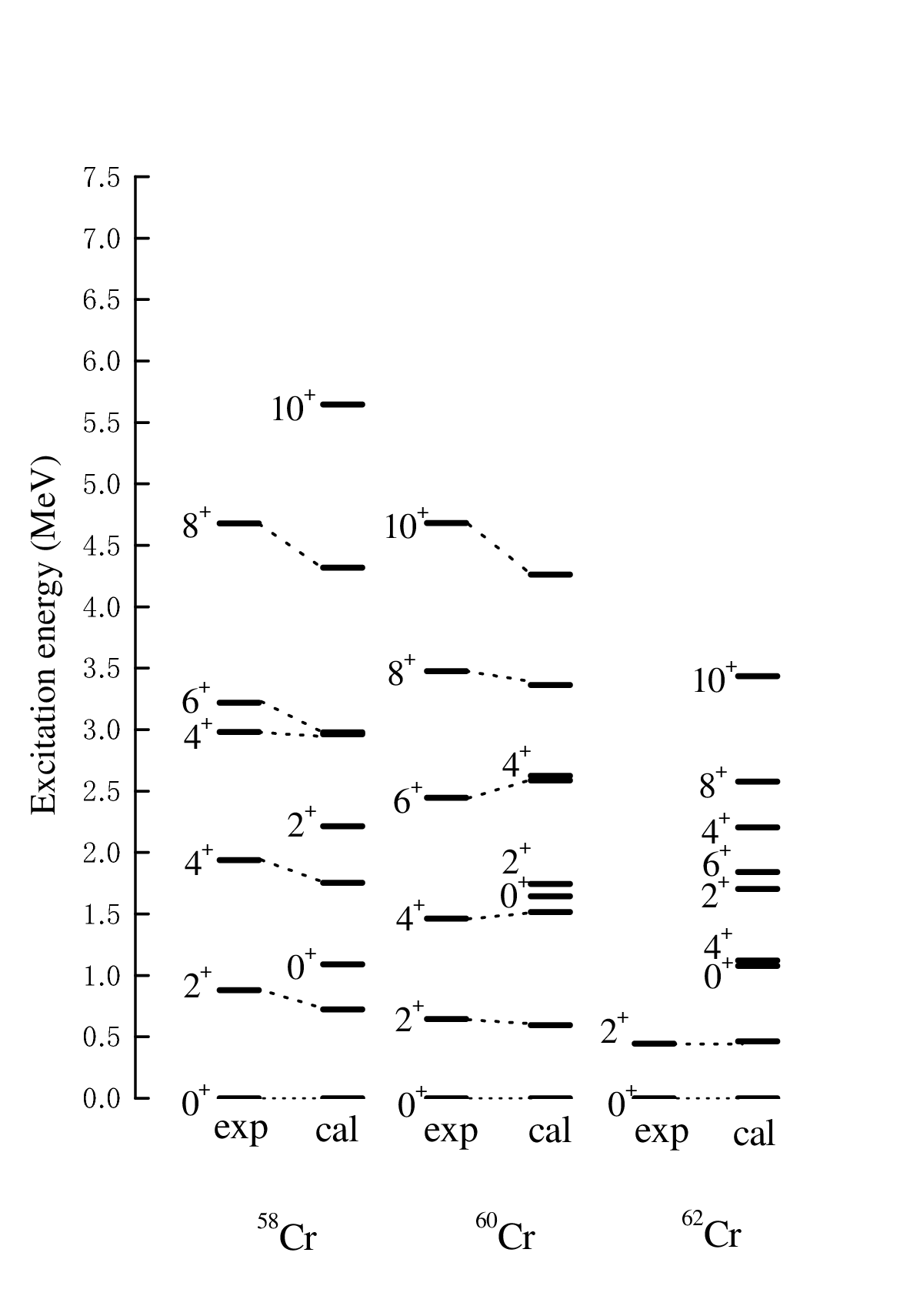}
\protect\caption{\label{fig1} Experimental and calculated energy
levels of positive-parity states for even-even isotopes
$^{52-62}$Cr.}
\end{figure*}
\begin{figure}[t]
\includegraphics[totalheight=9.5cm]{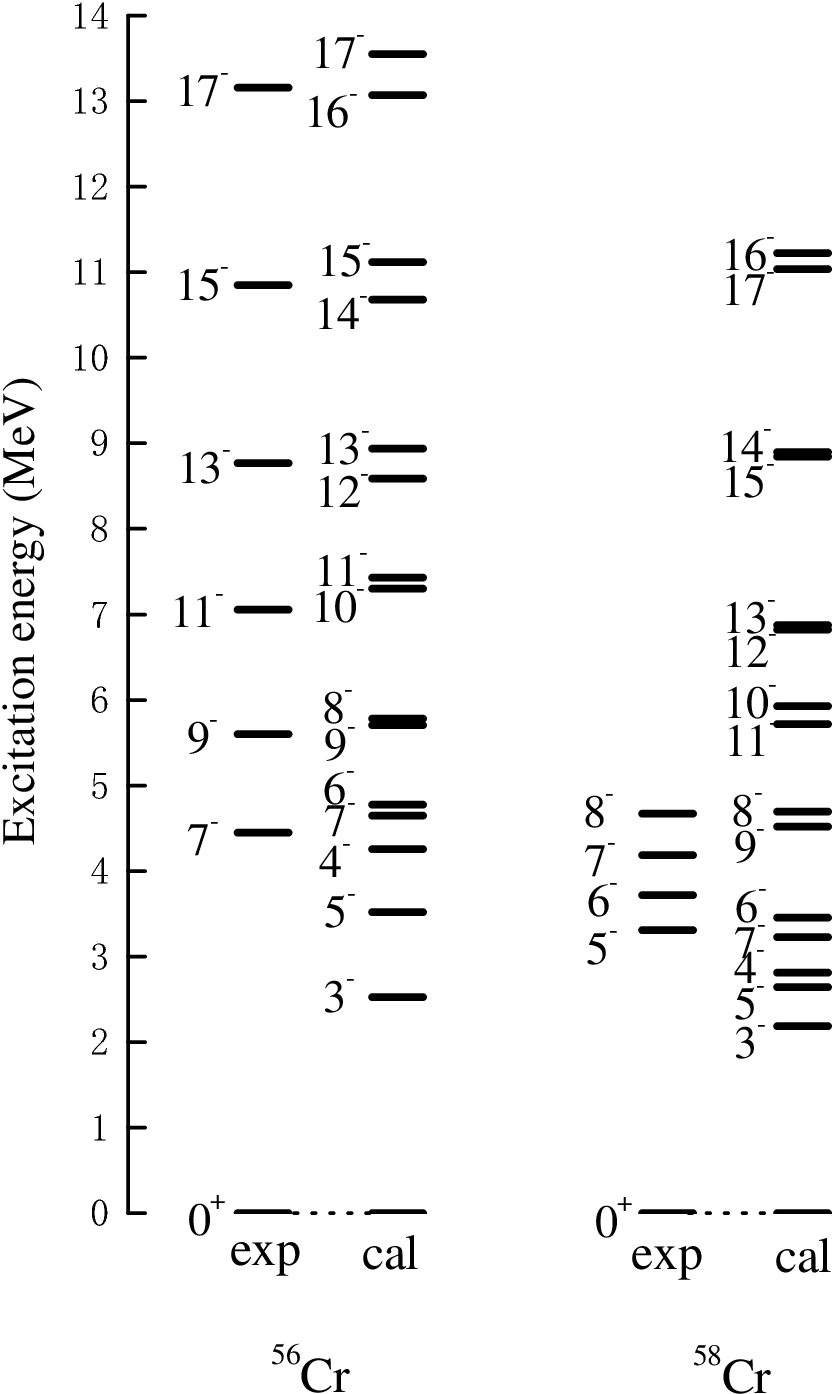}
\caption{ Experimental and calculated energy levels of
negative-parity states for $^{56,58}$Cr.}
  \label{fig2}
\end{figure}

\section{The shell model}\label{sec2}

We start with the following form of Hamiltonian, which consists of
pairing plus multipole terms with the monopole interaction included:
\begin{eqnarray}
 H & = & H_{\rm sp} + H_{P_0} + H_{P_2} + H_{QQ} + H_{OO}
       + H^{T=0}_{\pi \nu} + H_{\rm mc}  \nonumber \\
   & = & \sum_{\alpha} \varepsilon_a c_\alpha^\dag c_\alpha
    -  \sum_{J=0,2} \frac{1}{2} g_J \sum_{M\kappa} P^\dag_{JM1\kappa} P_{JM1\kappa}  \nonumber \\
   & - & \frac{1}{2} \chi_2/b^{4} \sum_M :Q^\dag_{2M} Q_{2M}:
         - \frac{1}{2} \chi_3/b^{6} \sum_M :O^\dag_{3M} O_{3M}: \nonumber \\
   & - & k^0 \sum_{a \leq b} \sum_{JM} A^\dagger_{JM00}(ab) A_{JM00}(ab)
            \nonumber \\
   & + & \sum_{a \leq b} \sum_{T} k_{\rm mc}^T(ab) \sum_{JMK}
               A^\dagger_{JMTK}(ab) A_{JMTK}(ab),
          \label{eq:0}
\end{eqnarray}
where $b$ in the third and fourth terms is the length parameter of
harmonic oscillator. We take the $J=0$ and $J=2$ interactions in the
pairing channel, and the quadrupole-quadrupole ($QQ$) and
octupole-octupole ($OO$) terms in the particle-hole channel
\cite{Hasegawa01,Kaneko02}. The monopole interaction is divided into
two parts, namely the average $T=0$ monopole field $H^{T=0}_{\pi
\nu}$ and the monopole correction term $H_{\rm mc}$. The Hamiltonian
(\ref{eq:0}) is isospin invariant.

Our calculation has been performed by using the ANTOINE shell-model
code \cite{Antoine}. We consider the $^{48}$Ca core as mentioned
above, and employ the single-particle energies $\varepsilon_{f7/2}
=0.0$, $\varepsilon_{p3/2} =2.0$, $\varepsilon_{p1/2} =4.0$,
$\varepsilon_{f5/2} =5.5$ (all in MeV) in the calculation. The
relative single-particle energies of $f_{7/2}$, $p_{3/2}$,
$p_{1/2}$, and $f_{5/2}$ are taken from the excitation energies of
the low-lying negative-parity states in $^{49}$Ca, and the energy
between $\varepsilon_{f7/2}$ and $\varepsilon_{p3/2}$ is determined
from the excitation energy of 3/2$^{-}$ state in $^{41}$Ca. The
single-particle energy $\varepsilon_{g9/2} =4.5$ MeV is chosen so as
to reproduce the $9/2_{1}^{+}$ energy level in $^{53}$Cr. It is not
surprising that the $g_{9/2}$ orbit lies below the $f_{5/2}$ orbit
because the 9/2$^{+}$ level is lower than the 5/2$^{-}$ level in
$^{49}$Ca \cite{Firestone}. This lowering of the $g_{9/2}$ would be
attributed to the attractive $T=1$ monopole interaction
$V_{f_{7/2},g_{9/2}}^{T=1}$.

We adopt the following interaction strengths for the pairing plus
multipole forces
\begin{eqnarray}
 & {} &  g_0 = 17.89/A, \quad g_2 = 152.24/A^{5/3},            \nonumber \\
 & {} &  \chi_2 = 228.36/A^{5/3},
         \chi_3 = 485.1/A^2 ~(\mbox{in MeV}). \label{eq:1}
\end{eqnarray}
The average monopole force $H^{T=0}_{\pi\nu}$ is neglected because
it does not affect the low-lying excited states of neutron-rich Cr
isotopes. For the monopole correction terms, we use
\begin{eqnarray}
  & {} & k_{\rm mc}^{T=0}(f_{7/2},p_{3/2}) = 0.5,
   \quad k_{\rm mc}^{T=0}(f_{7/2},f_{5/2}) = -0.6,  \nonumber \\
  & {} & k_{\rm mc}^{T=0}(f_{7/2},p_{1/2}) = 0.4,
   \quad k_{\rm mc}^{T=1}(p_{3/2},g_{9/2}) = -1.0.
\label{eq:2}
\end{eqnarray}
The repulsive $\pi f_{7/2}$-$\nu p_{3/2}$ and $\pi f_{7/2}$-$\nu
p_{1/2}$ monopole interactions can give rise to the correct order
of energy levels in odd-mass Cr isotopes, and also lower the first
excited $0_{2}^{+}$ state for the even-even Cr nuclei near $N=28$.
From comparison of the low-lying levels of $^{49}$Ca with those of
$^{57}$Ni, one sees clearly that the $5/2_{1}^{-}$ state in
$^{57}$Ni requires a monopole interaction between $\pi f_{7/2}$
and $\nu f_{5/2}$. This interaction
also plays an important role in lowering the first excited
$0_{2}^{+}$ state around $N=40$. The sharp decrease \cite{Deacon}
of the low-lying $9/2_{1}^{+}$ state observed along the isotopic
chain from $^{53}$Cr to $^{59}$Cr is considered as a clear
indication for the monopole interaction, and the attraction
between the $\nu p_{3/2}$ and $\nu g_{9/2}$ orbitals pushes down
the $9/2_{1}^{+}$ state. Having considered the above monopole
interactions, we choose the single-particle energy of the
$g_{9/2}$ orbital so as to reproduce the $9/2_{1}^{+}$ energy in
$^{53}$Cr. For calculations of spectroscopic $Q$-moments and
$B(E2)$ values, we take the standard effective charges $e_\pi=1.5e$
for protons and $e_\nu=0.5e$ for neutrons.

To get insight of the characteristic behavior of these neutron-rich
Cr isotopes, we further consider mean-field approximations for our
shell model Hamiltonian. The so-obtained effective single-particles
are the dressed particles that carry information on the
interactions. In Fig. \ref{fig11}, we present the calculated neutron
effective single-particle energies (SPE) \cite{Kaneko06}. The upper
(lower) panel in Fig. \ref{fig11} shows the effective SPE in the
shell-model Hamiltonian (\ref{eq:0}) with (without) the monopole
interactions (\ref{eq:2}). In the upper panel, we can see drastic
variations of the effective SPE for each orbitals.

\begin{figure*}[htb]
\includegraphics[width=0.65\textwidth]{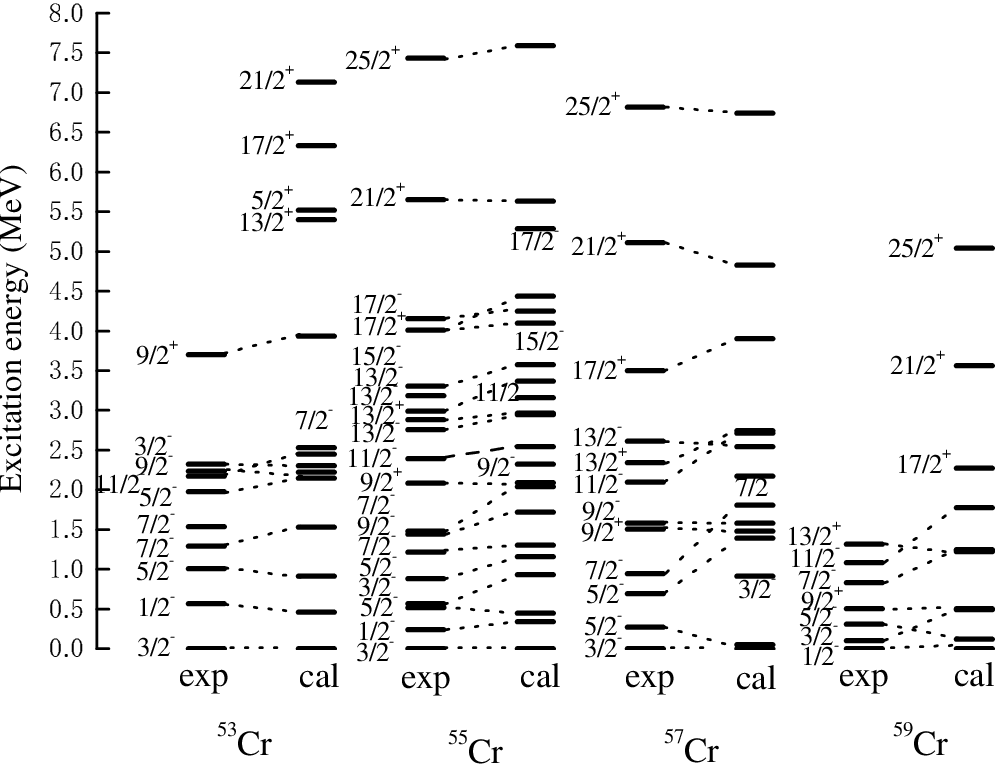}
\protect\caption{ Experimental and calculated
energy levels for the odd-mass $^{53-59}$Cr.} \label{fig3}
\end{figure*}

\begin{figure}[t]
\includegraphics[totalheight=5.5cm]{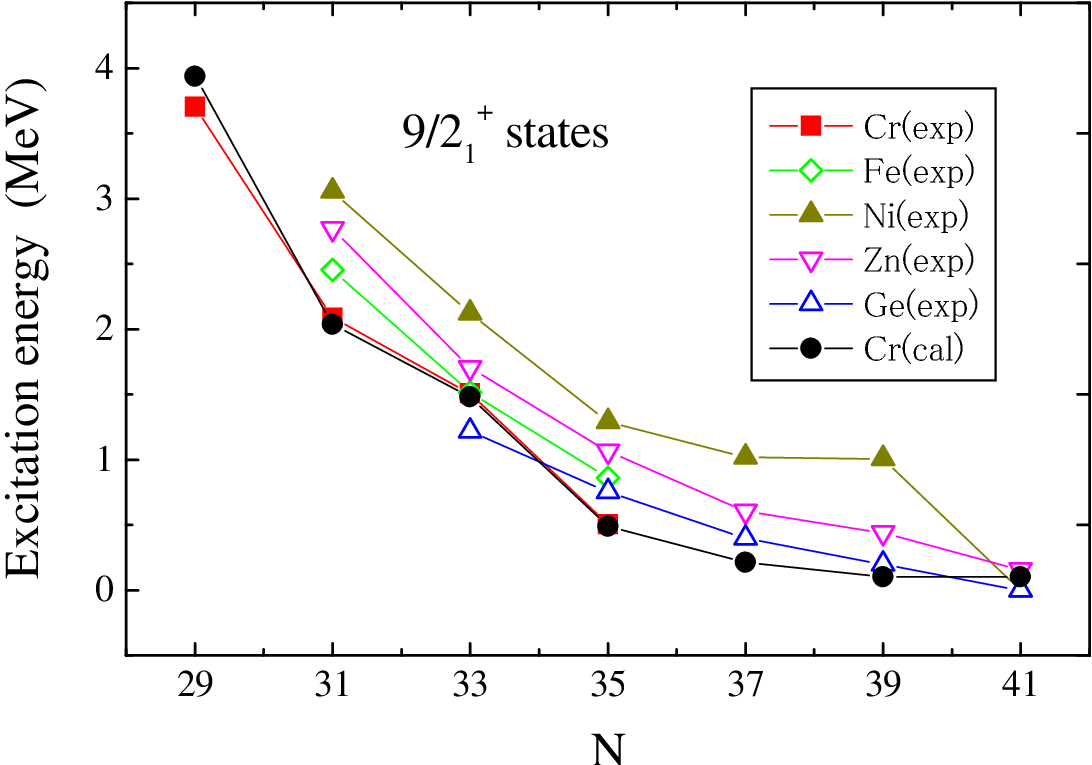}
\caption{ (Color online) Systematics of the first excited
$9/2_{1}^{+}$ states for Cr, Fe, Ni, Zn, and Ge isotopes. The
calculated energy levels are compared with experimental data for
odd-mass Cr nuclei.}
  \label{fig4}
\end{figure}

First, starting from $N=28$, the effective SPE for the $g_{9/2}$
orbital decreases quickly with increasing neutron number, and then
becomes degenerate with the $p_{3/2}$ orbital in the range from
$N=32$ to $N=40$. The sudden drop of the first excited $9/2_{1}^{+}$
in the odd-mass nuclei $^{53-59}$Cr (see later in Figs. \ref{fig3}
and \ref{fig4}) is related to this behavior of $g_{9/2}$, while the
pairing correlation increases the excitation energy of $9/2_{1}^{+}$
by about 1 MeV. This rapid decrease of the effective SPE is
attributed to the strong attractive monopole interaction between the
$\nu p_{3/2}$ and $\nu g_{9/2}$ orbitals in Eq. (\ref{eq:2}).

Second, we can see a change of the SPE level spacings among the
orbitals around $N=40$. This change is also due to the monopole
interactions in Eq. (\ref{eq:2}). The repulsive $\pi f_{7/2}$-$\nu
p_{3/2}$ and $\pi f_{7/2}$-$\nu p_{1/2}$ monopole interactions give
a correct sequence of the energy levels in odd-mass Cr isotopes. The
attractive $\pi f_{7/2}$-$\nu f_{5/2}$ monopole interaction plays an
important role in lowering the first excited $0_{2}^{+}$ state
around $N=40$. As already mentioned before, the attraction between
the $\nu p_{3/2}$ and $\nu g_{9/2}$ orbitals pushes down the
$9/2_{1}^{+}$ state, and provides an explanation for the sharp
decrease of the low-lying $9/2_{1}^{+}$ state from $^{53}$Cr to
$^{59}$Cr. These monopole interactions cooperatively give rise to
the reduction of the effective SPE spacing. The deformation that
develops around $N=40$ is due to the quasi-degeneracy of the
effective SPE spacing and the collectivity of the intruder $\nu
g_{9/2}$ orbital.


\section{Results and discussions}\label{sec3}

\subsection{Energy levels}

Let us first present the shell model results for even-even Cr
isotopes. Figure \ref{fig1} shows a comparison between calculated
results for positive-parity states and the experimental data
\cite{Zhu,Firestone} for the even-even Cr isotopes $^{52-62}$Cr. As
one can see, the theoretical results describe satisfactorily the
experimental energy levels. The feature that the first excited
$2_{1}^{+}$ energy is steadily decreasing as neutron number
approaching $N=40$ is correctly reproduced. One can also see a local
increase of $2_{1}^{+}$ energy at $^{56}$Cr.

The calculated first excited $0_{2}^{+}$ states for $^{52-56}$Cr
show a decreasing trend with increasing neutron number, although the
calculated $0_{2}^{+}$ energy for $^{52}$Cr lies too high as
compared to experiment. This discrepancy may be caused by the
assumption of the $N=28$ shell closure in the present calculation.
The neutron excitations of the $N=28$ core, which are neglected in
the present study, would lower the $0_{2}^{+}$ state in $^{52}$Cr.
The situation is quite similar to the anomalous behavior of the
$0_{2}^{+}$ state in $^{54}$Fe. In fact, the full $fp$-shell model
calculation \cite{Caurier1} has shown that the $0_{2}^{+}$ states in
$^{52}$Cr and $^{54}$Fe are the 2p-2h band-head. The present
calculation predicts the $0_{2}^{+}$ energy levels in $^{58-62}$Cr
which have not yet been observed. It is interesting to see a
particular prediction that the calculated $0_{2}^{+}$ level jumps up
at $^{60}$Cr with $N=36$ while the $2_{1}^{+}$ level in the same
nucleus decreases. The attractive $\nu p_{3/2}$-$\nu g_{9/2}$
monopole interaction pushes the $g_{9/2}$ orbital down, causing the
single-particle spacing between the $g_{9/2}$ and the $p_{1/2}$
orbital to increase.

Figure \ref{fig2} shows a comparison between the calculated and
experimental energy levels for the negative-parity states in
$^{56}$Cr and $^{58}$Cr \cite{Zhu}. In $^{56}$Cr, the odd-spin
levels of $7_{1}^{-}$, $9_{1}^{-}$, $11_{1}^{-}$, $13_{1}^{-}$,
$15_{1}^{-}$, and $17_{1}^{-}$ are reproduced very well, while the
even-spin levels are predicted. The $3_{1}^{-}$ states in $^{56}$Cr
and $^{58}$Cr are predicted to lie around 2.5 MeV. The $3_{1}^{-}$
state is considered as an 1p-1h excitation from the negative-parity
orbital $p_{3/2}$ to the positive parity orbital $g_{9/2}$ because
the $3_{1}^{-}$ energies are comparable to the single-particle
spacing 2.5 MeV between these orbitals. In addition, it is
interesting to see a predicted formation of spin doublet states in
both nuclei, as for instance the pairs $10_{1}^{-}$-$11_{1}^{-}$,
$12_{1}^{-}$-$13_{1}^{-}$, $14_{1}^{-}$-$15_{1}^{-}$, and
$16_{1}^{-}$-$17_{1}^{-}$. In $^{58}$Cr, the even-spin levels of
$6^{-}$ and $8^{-}$ are reproduced well, although the calculated
odd-spin $5_{1}^{-}$ and $7_{1}^{-}$ levels are lower than the
experimental ones.

In Fig. \ref{fig3}, calculated results for the odd-mass Cr isotopes
$^{53-59}$Cr are shown, and compared with experimental data
\cite{Zhu,Firestone}. Overall, the calculation reproduces the data
well. In particular, the agreement for $^{53,55}$Cr with both
negative- and positive-parity states is excellent. In $^{57}$Cr,
most energy levels are described well; only the low-lying
$5/2_{2}^{+}$ and $7/2_{1}^{+}$ states are calculated higher than
the experimental ones. In these odd-mass Cr isotopes, the most
striking feature is that the first excited $9/2_{1}^{+}$ state drops
down rapidly when neutrons are added: from 2087 kev in $^{55}$Cr to
1057 kev in $^{57}$Cr, and down to 503 kev in $^{59}$Cr. The
calculation reproduces this systematics very well. This rapid
decrease of the $9/2_{1}^{+}$ excitation energy can be understood as
due to the attractive monopole interaction between the $\nu g_{9/2}$
and $\nu p_{3/2}$ orbitals that pushes down the $\nu g_{9/2}$
orbital. The high-spin states with positive-parity in $^{55,57}$Cr
are also reproduced well. In both nuclei, the rotational bands built
on the $9/2_{1}^{+}$ state provide a strong experimental evidence
for the shape-driving potential of this intruder level
\cite{Freeman,Deacon}. In fact, we can see large $E2$ transitions in
this band in later discussions.

At this point, it is worthwhile discussing systematics of the
$9/2_{1}^{+}$ state. In Fig. \ref{fig4}, the $9/2_{1}^{+}$ states of
the odd-mass Cr are compared with those of Fe, Ni, Zn, and Ge
isotopes \cite{Firestone}. As already seen from Fig. \ref{fig3}, the
calculated energy levels of $9/2_{1}^{+}$ are in a good agreement
with experiment for $^{53-59}$Cr. Now in Fig. \ref{fig4}, all the
curves show that the $9/2_{1}^{+}$ state energy steadily decreases
with increasing neutron number. It should be pointed out that the
more rapid drop of the $9/2_{1}^{+}$ energy from $^{57}$Cr to
$^{59}$Cr correlates with the sudden increase of the calculated
$0_{2}^{+}$ energy from $^{58}$Cr to $^{60}$Cr in Fig. \ref{fig1}.
Further calculations beyond $N=37$ predict a low-excited
$9/2_{1}^{+}$ for $^{61-65}$Cr. Interestingly enough, the curve for
Cr isotopes is found to be very close to that of Ge isotopes,
namely, the odd-mass Cr and Ge isotones have very similar
$9/2_{1}^{+}$ energy. At $N=41$, the $9/2^{+}$ state becomes the
ground state for $^{69}$Ni and$^{73}$Ge, while it is an excited
state at 0.158 MeV for $^{71}$Zn. The calculation predicts that the
$9/2_{1}^{+}$ level is an excited state for $^{65}$Cr.

\begin{figure}[t]
\includegraphics[totalheight=9.5cm]{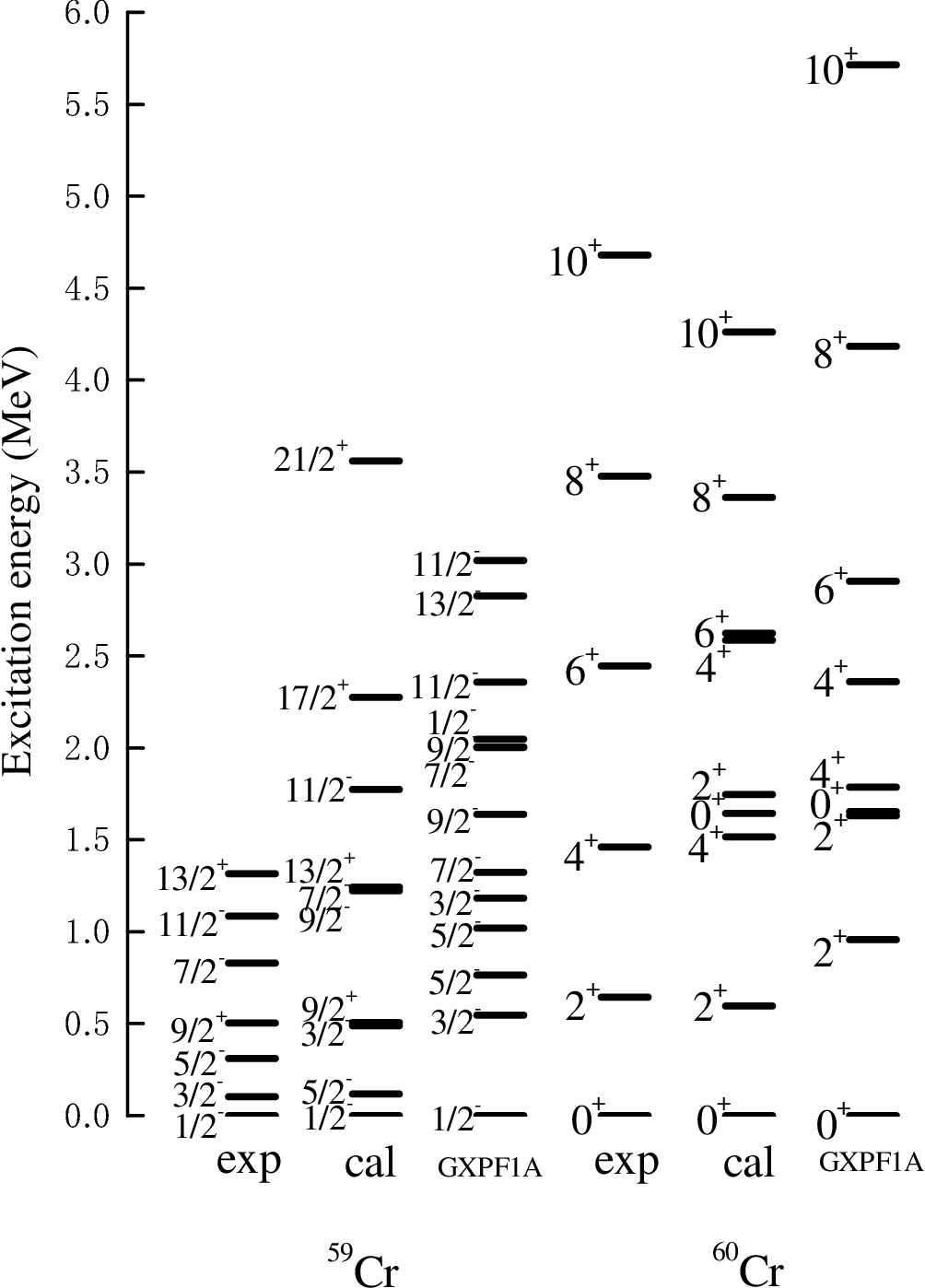}
\caption{ Comparison of experimental energy levels
with the present and the GXPF1A calculations for $^{59,60}$Cr.}
 \label{fig5}
\end{figure}

In Fig. \ref{fig5}, we compare the experimental energy levels of
$^{59}$Cr and$^{60}$Cr with the present and the GXPF1A calculation.
Our results in Fig. \ref{fig5}, as well as in the early discussed
Fig. \ref{fig1}, show a much better agreement. The present
calculation reproduces the positive-parity $9/2_{1}^{+}$ and
$13/2_{1}^{+}$ states in $^{59}$Cr, and predicts higher-spin states
above them. For $^{60}$Cr, our calculation nicely reproduces the
experimental high-spin levels as well. It is thus clear that the
$\nu g_{9/2}$ excitations play a significant role in the current
problem, and it is crucial to include this orbital for a correct
description of these nuclei.

\begin{figure*}[htb]
\includegraphics[width=0.65\textwidth]{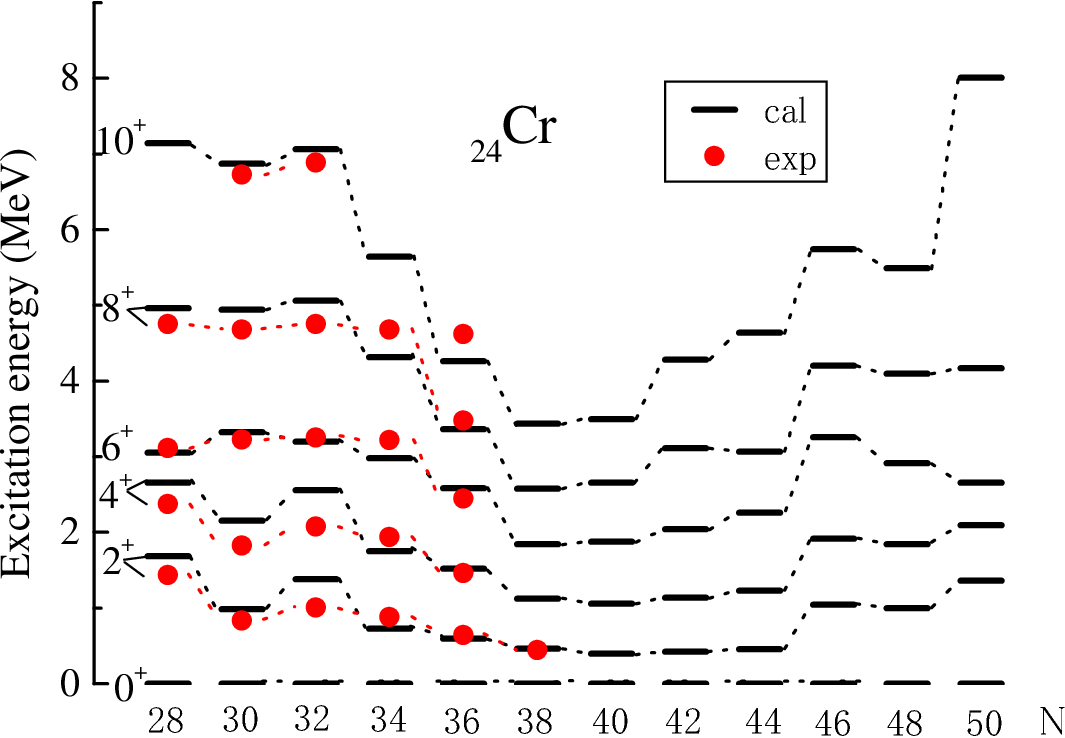}
\protect\caption{(color online) Experimental and calculated energy
levels of the yrast states for the even-even Cr isotopes with
$N=28-50$.}
 \label{fig6}
\end{figure*}
\begin{figure}[t]
\includegraphics[totalheight=6.5cm]{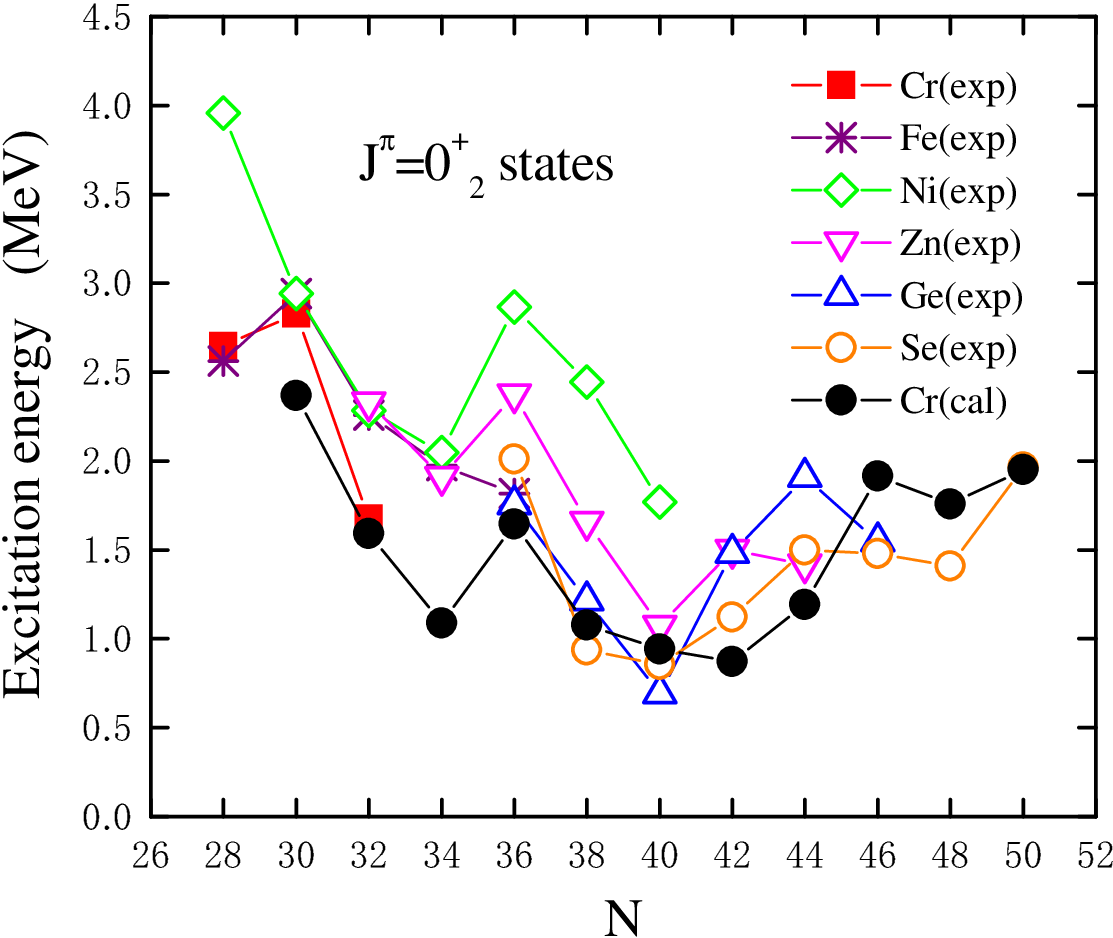}
\caption{(color online) Systematics of the first excited
$0_{2}^{+}$ states for Cr, Fe, Ni, Zn, Ge, and Se isotopes. The
calculated energy levels are compared with the experimental data
for Cr isotopes.}
  \label{fig7}
\end{figure}

The above success motivates us to move in on the even more
neutron-rich region, where no experimental data are available. Let
us next investigate the Cr isotopes beyond $N=38$. Recent
observation of very low first excited $2^{+}$ states in $^{60}$Cr
and $^{62}$Cr suggests that these nuclei are deformed
\cite{Sorlin02}. We thus expect that collective motion may be the
dominant mode for the low-lying states in the isotopes around
$N=40$. Figure \ref{fig6} shows the calculated energy levels of
yrast states for the neutron-rich even-even isotopes $^{52-74}$Cr.
For those below $N=36$, the calculated results reproduce well the
observed trend of experimental data. In $^{56}$Cr, the calculation
correctly reproduces the enhanced excitation energy of the first
excited $2_{1}^{+}$ and $4_{1}^{+}$ state as compared with the
corresponding states in the neighboring $^{54,58}$Cr, which is, as
discussed early, a strong signature of the subshell gap at $N=32$.

There is a striking behavior along the isotopic chain: the abrupt
change in structure at $N=36$ in $^{60}$Cr. As can be clearly seen
from Fig. \ref{fig6}, the $6_{1}^{+}$, $8_{1}^{+}$, and $10_{1}^{+}$
state energies drop down suddenly at this neutron number. Another
notable structure change, as shown in Fig. \ref{fig7}, is that the
excitation energy of the $0_{2}^{+}$ state exhibits a peak at
$N=36$. Remarkably, a similar structure evolution has been seen
along the $N=Z$ line in the proton-rich region, and a abrupt change
in structure in the $N=Z=36$ nucleus $^{72}$Kr has been discussed
\cite{Kr72} by the present authors. In Ref. \cite{Kr72}, it was
discussed that the yrast energy levels in the $N=Z < 36$ nuclei
$^{64}$Ge and $^{68}$Se show larger separations, corresponding to
smaller moments of inertia, but in $^{72}$Kr, the level separation
in the yrast sequence goes down drastically. We have suggested in
Ref. \cite{Kr72} that this is an example of phase transition along
the $N=Z$ line.

We now discuss the isotopes beyond $N=36$. In the $N=38$ isotope
$^{62}$Cr, only the first excited $2_{1}^{+}$ state has been
observed \cite{Sorlin03}. Together with this state, a general trend
starting from $N=32$ can be clearly seen, in which the $2_{1}^{+}$
excitation energies decrease gradually with increasing neutron
number. The calculated excitation energies of yrast states further
drop down at the $N=38$ and 40 isotopes. The abrupt changes in
structure across neutron-rich Cr isotopes appear as a result of low
single-particle level density, which leads to a well-developed
deformation at $N=40$. When going toward more neutron-rich region
from $N=40$ excitation energies of the yrast states begin to
increase with increasing neutron number, which makes the $N=40$
isotope $^{64}$Cr the most deformed among the isotopic chain. It was
proposed by Zuker {\it et al.} \cite{Zuker} that a minimal valence
space to be able to develop quadrupole collectivity should contain
at least a $(j,j-2,....)$ sequence of orbits. In the upper
$g_{9/2}$-shell nuclei, the $g_{9/2}$ orbital and its quasi-SU(3)
counterpart the $d_{5/2}$ orbital have to be taken into account in
order to reproduce experimental data. Therefore, to describe the
neutron upper $g_{9/2}$-shell nuclei, it would be better to take
into account the $d_{5/2}$ orbital. At present, however, it is not
possible to include the $d_{5/2}$ orbital to the shell model space
because of huge dimensions of the configuration space.

In Fig. \ref{fig7}, we show the calculated and experimental
$0_{2}^{+}$ states for even-even Cr isotopes, and for comparison,
the experimentally known $0_{2}^{+}$ states for even-even Fe, Ni,
Zn, Ge, and Se isotopes \cite{Firestone} are also plotted. For the
only three Cr isotopes for which data exist, the calculation
reproduces the data of $^{54,56}$Cr. The calculated $0_{2}^{+}$
level at 5.6 MeV for $^{52}$Cr (not shown in Fig. \ref{fig7}) lies
too high when compared with the experimental one. The reason for
this discrepancy may be attributed to the fact that neutrons in the
$f_{7/2}$ orbital are frozen in the calculation. From the
systematics in Fig. \ref{fig7}, we can see two characteristics for
these first excited $0_{2}^{+}$ levels. One is the presence of a
peak at $N=36$ in the experimental data for Ni and Zn isotopes as
well as in the calculated results for Cr isotopes. For Ge and Se
isotopes, $N=36$ might also show a $0_{2}^{+}$ peak; however, it is
not conclusive at present because there are no data for the
$0_{2}^{+}$ energy at $N=34$. The other characteristic in Fig.
\ref{fig7} is the occurrence of a minimum in the $0_{2}^{+}$ energy
level around $N=40$. This minimum occurs because of the monopole
interaction between $\pi f_{7/2}$ and $\nu f_{5/2}$.

\begin{figure}[t]
\includegraphics[totalheight=9.5cm]{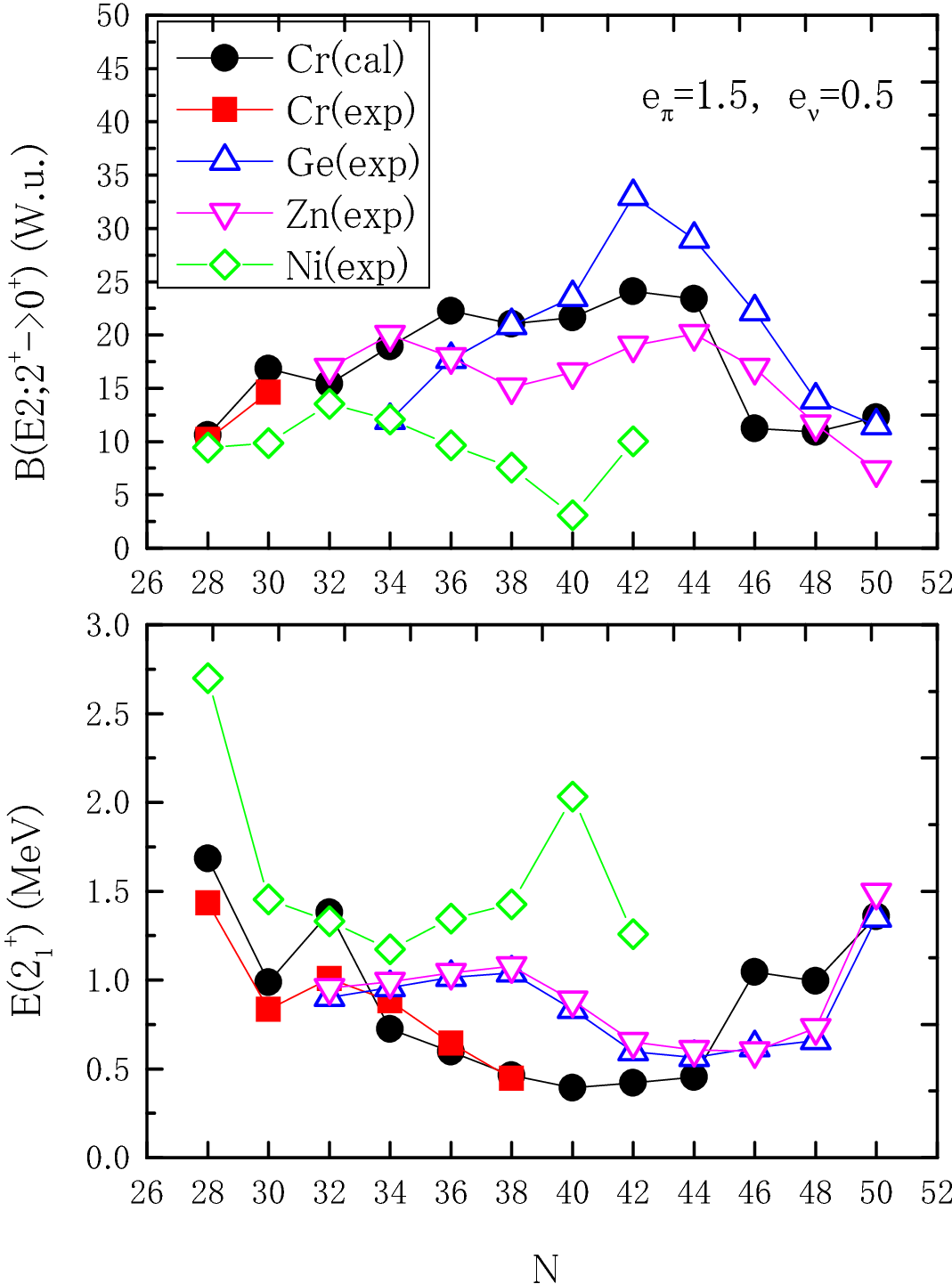}
\caption{(color online) Calculated and experimental values of
(upper panel) $B(E2,0_{1}^{+}\rightarrow 2_{1}^{+})$, and (lower
panel) $E(2_{1}^{+})$ for even-even Cr, Ni, Zn, and Ge isotopes. }
  \label{fig8}
\end{figure}
\begin{figure}[t]
\includegraphics[totalheight=5.5cm]{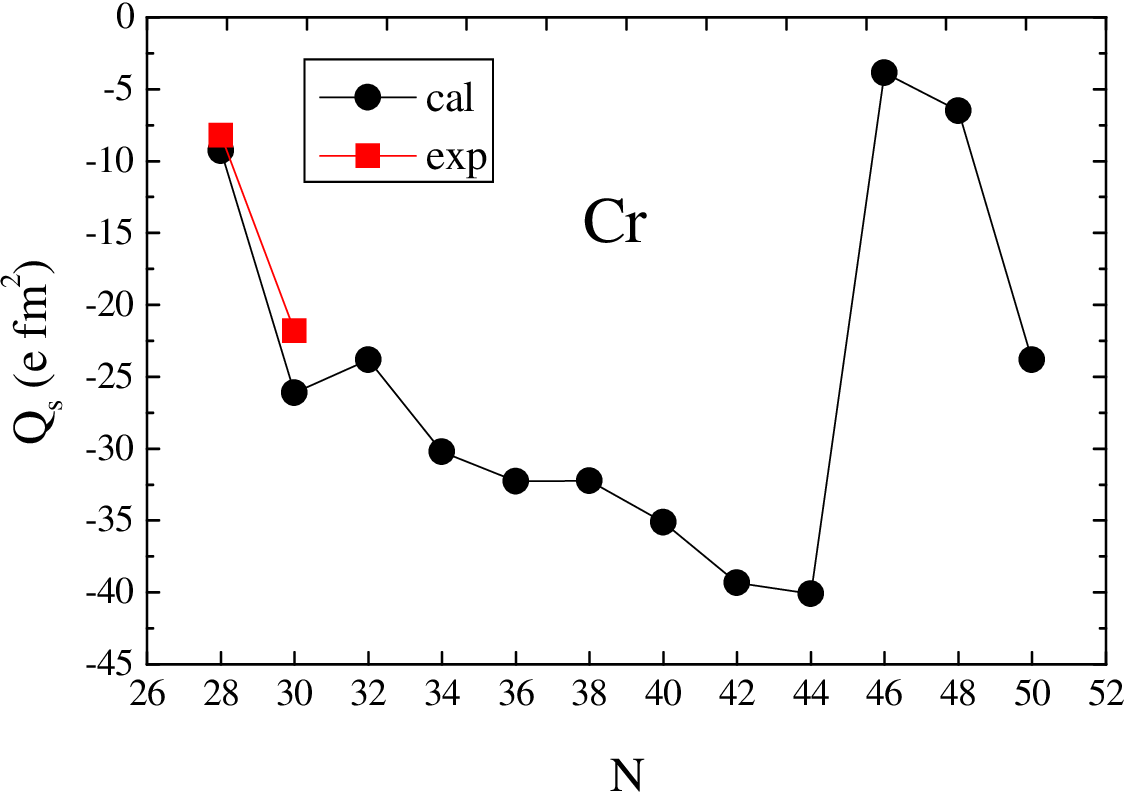}
\caption{(color online) Calculated and experimental spectroscopic
qudrupole moment for even-even Cr isotopes. }
  \label{fig9}
\end{figure}

\subsection{$B(E2)$ values}

Calculation of electromagnetic transition probabilities serves as a
strict test for wave functions of theoretical models. In the upper
panel of Fig. \ref{fig8}, experimental $B(E2;2_{1}^{+}\rightarrow
0_{1}^{+})$ dada for even-even Cr, Ni, Zn, and Ge isotopes are
collected \cite{Padilla,Walle,Firestone}, and our shell-model
calculations for Cr isotopes are compared with the known data. In
lower panel of Fig. \ref{fig8}, excitation energies of the first
$2^+$ state $E(2_{1}^{+})$ are plotted. With the usual effective
charges $e_{\pi}$ = 1.5$e$ and $e_{\nu}$ = 0.5$e$, the present
calculation including the $\pi f_{7/2}$ orbital in the model space
is found in good agreement with data for $^{52,54}$Cr. A sharp
contrasting behavior between Ni and other isotopes can be clearly
seen around $N=40$. The unusual $B(E2)$ and $E(2_{1}^{+})$ behavior
near $^{68}$Ni shows up at the subshell closure $N=40$. The $B(E2)$
values around $N=40$ are determined by filling the unique parity
$\nu g_{9/2}$ orbital and the proton core polarization. The present
calculation for Cr isotopes suggests that the predicted trend in
$B(E2)$ around $N=40$ is quite similar to the experimentally known
systematics in Zn and Ge isotopes \cite{Padilla,Walle}.

For shell model calculations in the model space consisting of the
$1p_{3/2}$, $0f_{5/2}$, $1p_{1/2}$, $0g_{9/2}$ orbitals for protons
and neutrons with an inert $^{56}$Ni core, large effective charges
$e_{\pi}$ = 1.9$e$ and $e_{\nu}$ = 0.9$e$ were used to obtain the
experimental $B(E2)$ values for Zn and Ge isotopes
\cite{Padilla,Walle,Hasegawa07}. This implies that excitation from
the $\pi f_{7/2}$ orbital would be important for obtaining large
$B(E2)$ values around $N=40$ \cite{Sorlin02}. It should be noted
that the calculated $E2$ transitions for $^{60,62,64}$Cr yield
nearly the same values as the recent shell model calculations in the
$fpgd$ space including the $d_{5/2}$ orbital \cite{Sorlin03}, where
their $B(E2)$ values are, respectively, 20.64, 20.7, 20.9 W.u. for
$^{60,62,64}$Cr. The calculated $2_{1}^{+}$ energies in Ref.
\cite{Sorlin03} diverge from the experimental trend as they remain
constant up to $N=38$ and increase at $^{64}$Cr.

The deformation estimated from the calculated
$B(E2;2_{1}^{+}\rightarrow 0_{1}^{+})$ around $N=40$ is
$\beta_{2}\sim$ 0.3, consistent with those obtained from the
empirical formula \cite{Sorlin03} and the Skyrme HFB calculations
\cite{Stoitsov03,Stoitsov05}. In the lower panel of Fig. \ref{fig8},
we can see the $N=32$ gap in the Cr isotopes. With increasing
neutron number beyond $N=32$, the first excited $2_{1}^{+}$ states
decrease steadily toward $N=40$. The $2_{1}^{+}$ energy systematics
of Cr behaves quite differently from Zn and Ge isotopes. The sudden
drop in $B(E2)$ values and large $2_{1}^{+}$ excitation energies for
$^{70-74}$Cr may be due to the neglect of the $d_{5/2}$ orbital
\cite{Sorlin02} in the present model space.

Figure \ref{fig9} shows a comparison for the calculated and
experimental spectroscopic quadrupole moments $Q_{s}$. The
calculated $Q_{s}$ values for $^{52,54}$Cr agree well with the data.
The drastic variation in the calculated $Q_{s}$ values for
$^{70-74}$Cr may also be an artefact due to the neglect of the
$d_{5/2}$ orbital, which causes variations in the $B(E2)$ values and
$2_{1}^{+}$ energy systematics in these nuclei, as seen in Fig.
\ref{fig8}.

\begin{figure}[t]
\includegraphics[totalheight=6.5cm]{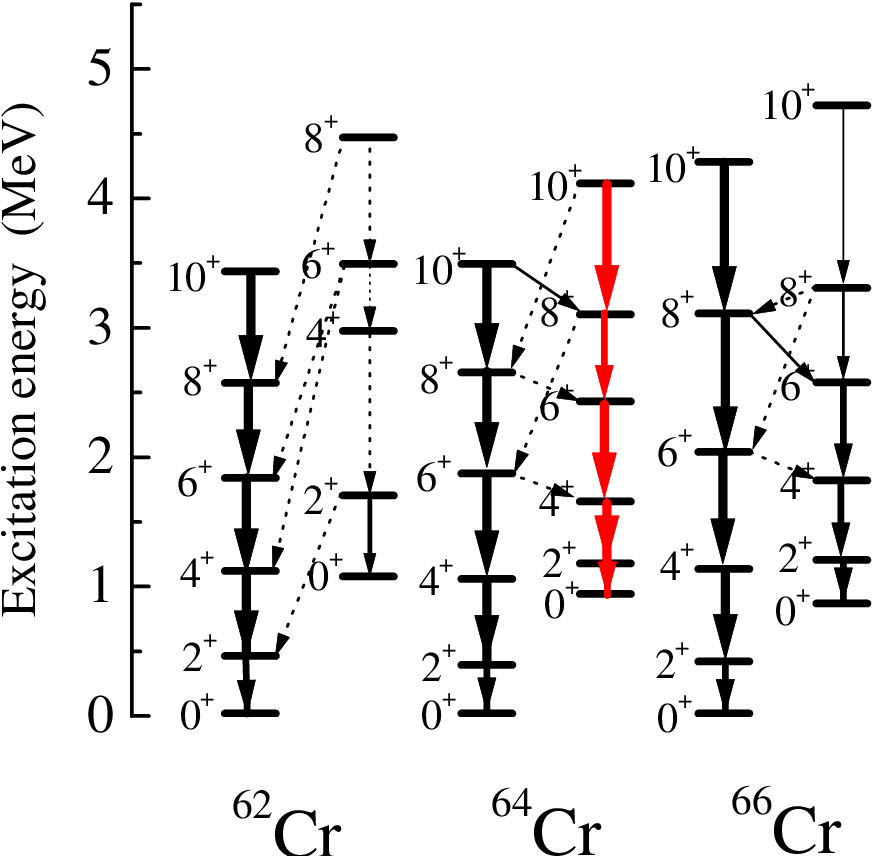}
\caption{(color online) Calculated level scheme for
$^{62,64,66}$Cr. The widths of the arrows denote relative values
of $B(E2)$.}
  \label{fig10}
\end{figure}

In a previous paper \cite{Kaneko06}, we predicted a new band built
on the first excited $0_{2}^{+}$ state in two neutron-rich nuclei
$^{68}$Ni and $^{90}$Zn. The structure of states in this band is
dominated by 2p-2h excitations from the $fp$ shell to the intruder
$g_{9/2}$ orbital. We expect that an analogous band exists also in
$^{64}$Cr with $N=40$. The structure of this anticipated band is
quite different from that of the ground state. This happens because
the opposite signs of parity between the $g_{9/2}$ orbital and the
$fp$ shell do not favor 1p-1h excitations. Figure \ref{fig10} shows
the theoretical level schemes. The predicted $B(E2)$ values for
$^{62}$Cr and $^{64}$Cr are summarized in Table I. One can see
strong enhancement in $B(E2)$ values in the side band built on the
first excited $0_{2}^{+}$ state in $^{64}$Cr. The
$B(E2;2_{1}^{+}\rightarrow 0_{1}^{+})$ and
$B(E2;4_{1}^{+}\rightarrow 2_{1}^{+})$ values for $^{62,64}$Cr are
consistent with the calculated values by Sorlin {\it et al.}
\cite{Sorlin02}. This situation is very similar to that in the
neutron-rich $^{68}$Ni and $^{90}$Zn. The relevant ingredient here
is excitations from the $fp$ shell to the intruder $g_{9/2}$
orbital, and we expect that this is a common feature for the
isotones of $N=40$. In Table II, $B(E2)$ values for the
positive-parity band built on the $9/2_{1}^{+}$ state in the
odd-mass nuclei, $^{53-59}$Cr, are shown. We can see that for
$^{59}$Cr $E2$ transition rates have large values, and therefore
show a strong collectivity. The deformation estimated from the
calculated results is $\beta_{2}=0.18$, consistent with the Total
Routhian Surface (TRS) calculations \cite{Deacon}.

\begin{table}[b]
\caption{Calculated $B(E2)$ values for the positive-parity yrast states and the
         excited states in $^{62}$Cr and $^{64}$Cr. These values are compared with
         those of Sorlin {\it et al.} \cite{Sorlin02}.}
\begin{tabular*}{85mm}{@{\extracolsep{\fill}}ccccc} \hline\hline
        & \multicolumn{2}{c}{$^{62}$Cr  [$e^2$fm$^{4}$]}
        & \multicolumn{2}{c}{$^{64}$Cr  [$e^2$fm$^{4}$]}    \\ \hline
$I_i^\pi \rightarrow I_f^\pi$ & Sorlin {\it et al.} & Calc. & Sorlin {\it et al.}  & Calc. \\ \hline
$2_1^+ \rightarrow 0_1^+$       &  302  &  307  &  318  &  329   \\
$4_1^+ \rightarrow 2_1^+$       &  428  &  436  &  471  &  460   \\
$6_1^+ \rightarrow 4_1^+$       &       &  449  &       &  464   \\
$8_1^+ \rightarrow 6_1^+$       &       &  455  &       &  421   \\
$10_1^+ \rightarrow 8_1^+$      &       &  500  &       &  523   \\
$2_2^+ \rightarrow 0_2^+$       &       &  243  &       &  293   \\
$4_2^+ \rightarrow 2_2^+$       &       &    5  &       &  407   \\
$6_2^+ \rightarrow 4_2^+$       &       &   17  &       &  381   \\
$8_2^+ \rightarrow 6_2^+$       &       &    7  &       &  322   \\
$10_2^+ \rightarrow 8_2^+$      &       &  335  &       &  371   \\ \hline\hline
\end{tabular*}
\label{table1}
\end{table}
\begin{table}[b]
\caption{Calculated $B(E2)$ values for the positive-parity band built on
         the $9/2_{1}^{+}$ state in the odd-mass nuclei $^{53-59}$Cr.}
\begin{tabular*}{85mm}{@{\extracolsep{\fill}}ccccc} \hline\hline
        & \multicolumn{2}{c}{$B(E2)$  [$e^2$fm$^{4}$]}    \\ \hline
$I_i^\pi \rightarrow I_f^\pi$ & $^{53}$Cr & $^{55}$Cr & $^{57}$Cr  & $^{59}$Cr \\ \hline
$13/2_1^+ \rightarrow 9/2_1^+$       &  143  &  283  &  260  &  360   \\
$17/2_1^+ \rightarrow 13/2_1^+$       &  123  &  300  &  280  &  406   \\
$21/2_1^+ \rightarrow 17/2_1^+$       &  58  &  172  &  126  &  361   \\
$25/2_1^+ \rightarrow 21/2_1^+$       &  73  &  148  &  114  &  305   \\ \hline\hline
\end{tabular*}
\label{table2}
\end{table}

\subsection{Ground-state energies and occupation numbers}

\begin{figure}[t]
\includegraphics[totalheight=5.5cm]{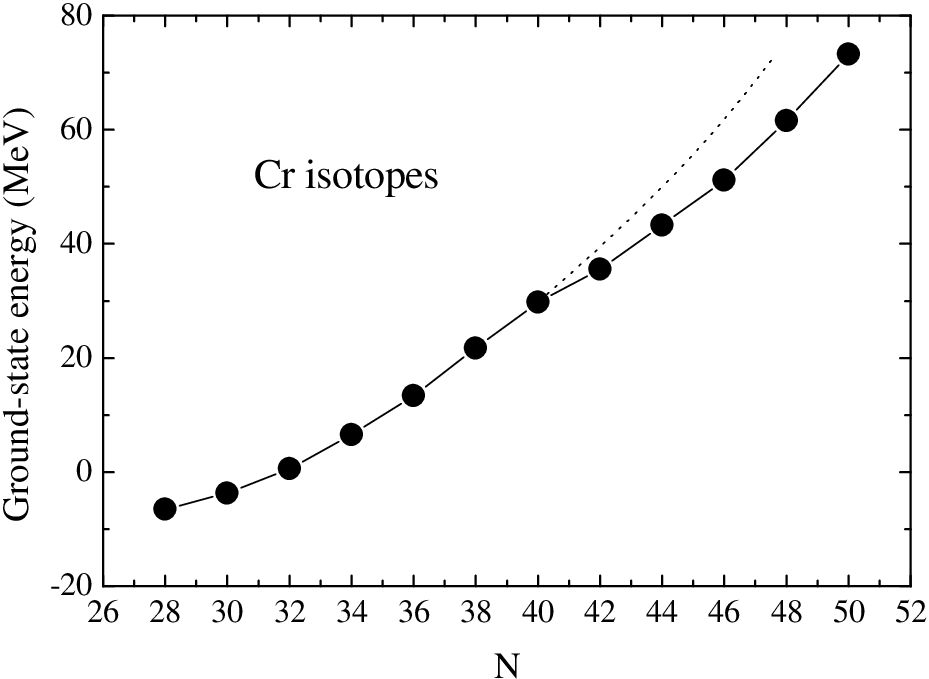}
\caption{Ground-state energies in the shell-model calculation for
even-even Cr isotopes. The dotted curve is an extrapolation
of the lower mass part of the results to show clearly the
discontinuity in the slope of the curve at $N=40$. }
  \label{fig12}
\end{figure}

It is striking that the effective SPE levels of the $\nu p_{3/2}$
and $\nu f_{5/2}$ orbitals decrease drastically with increasing
neutron number beyond $N=42$. This affects the properties of the
ground-state energy in the shell-model calculations. Indeed, as can
be clearly seen in Fig. \ref{fig12}, the ground-state energies bend
at $N=40$ so that it exhibits a discontinuity in the slope of the
curve in Fig. \ref{fig12}. This would correspond to a kind of phase
transition at $N=40$.

\begin{figure}[t]
\includegraphics[totalheight=9.5cm]{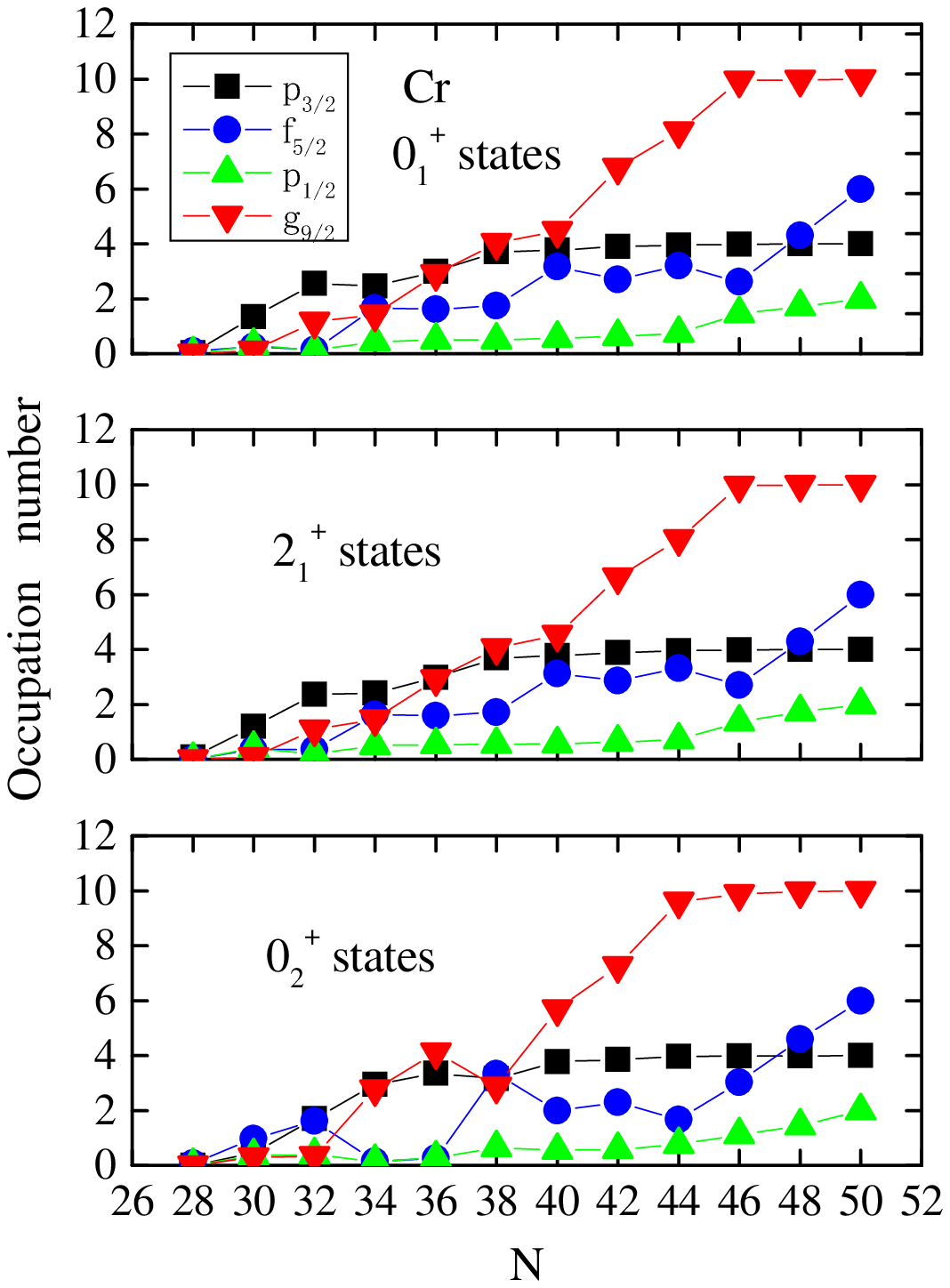}
\caption{(color online) Neutron occupation numbers of the
$fpg$-shell orbits for three low-lying states in even-even Cr
isotopes. (Upper panel) $0^+_1$ states; (middle panel) $2^+_1$
states; (lower panel) $0^+_2$ states.}
  \label{fig13}
\end{figure}

We can look at the above findings with other physical quantities
such as neutron occupation numbers of the $fpg$-shell orbits for the
low-lying $0_{1}^{+}$, $2_{1}^{+}$, and $0_{2}^{+}$ levels. The
upper panel in Fig. \ref{fig13} shows occupation numbers for the
ground states $0_{1}^{+}$. For nuclei with smaller $N$, the main
component is of the $\nu p_{3/2}$ orbital. At $N=36$ and 38, the
occupation number for the $\nu g_{9/2}$ orbital increases and nearly
equals to that of the $p_{3/2}$ orbital. At $N=40$, the three
occupation numbers of $\nu p_{3/2}$, $\nu g_{9/2}$, and $\nu
f_{5/2}$ take almost same values. Going to higher neutron numbers
from $N=40$, occupation number of the $\nu g_{9/2}$ orbital
increases drastically and becomes the dominant component. The
occupation numbers for the first excited $2_{1}^{+}$ states in the
middle panel show a similar distribution as in the ground states.
The bottom panel gives occupation numbers for the first excited
$0_{2}^{+}$ states. At $N=36$, we can see a peak of the $\nu
g_{9/2}$ orbital in occupation number, and the $\nu p_{3/2}$ and
$\nu g_{9/2}$ orbitals take almost the same occupation numbers. This
means that the $0_{2}^{+}$ state in $^{60}$Cr consists of
excitations from the $\nu p_{3/2}$ and $\nu g_{9/2}$ orbitals, and
lies higher than that of $N=34$. For $N=38$, the three orbitals
provide the collectivity. For the Cr isotopes beyond $N=40$, the
$\nu g_{9/2}$ orbital is the dominant component.

\section{Conclusions}\label{sec5}

We have carried out a comprehensive study for the structure of
neutron-rich Cr isotopes using the spherical shell model. Overall,
the calculated results are in good agreement with the experimentally
known energy levels for even-even $^{52-62}$Cr and odd-mass
$^{53-59}$Cr isotopes. Furthermore, a lowering of excitation
energies for neutron-rich Cr isotopes at and beyond $N=40$ has been
predicted. We have explained several characteristic behaviors found
in the calculation: (1) a rapid decrease of the $9/2_{1}^{+}$
energies for the odd-mass Cr isotopes with $N=29-35$, (2) drastic
structure changes at $N=36$ with enhancement of the $0_{2}^{+}$
energy and sudden reduction of the yrast band energy, and (3) steady
decrease of the $2_{1}^{+}$ energy in the smaller $N$ region and the
occurrence of pronounced collectivity around $N=40$. Several
interesting consequences associated with the subshell closure $N=40$
have been discussed: (1) enhanced $B(E2)$ values, (2) lowering of
the $0_{2}^{+}$ energies, and (3) a new band with strong
collectivity built on the $0_{2}^{+}$ state in $^{64}$Cr.

We have pointed out that the key ingredients to predict and explain
these striking features are the monopole interactions. These
attractive and repulsive interactions affect cooperatively the
effective SPE. As we have seen in Fig. \ref{fig11}, the effective
single-particle levels of 1$p_{3/2}$, 0$f_{5/2}$, and 0$g_{9/2}$
approach to each other near $N$ = 40, due to the monopole effects.
It seems that the cooperative effects of this behavior and the
many-body correlations enhance the collectivity near $N$ = 40. On
the other hand, the calculation without 1$d_{5/2}$ (see Refs.
\cite{Sorlin03} and \cite{Caurier}) does not lead to sufficient
collectivity, which may suggest a need of including the 1$d_{5/2}$
orbit in shell model calculations. It is an interesting open
question why the problem has not settled down and how one can unify
the different views. The attractive $\pi f_{7/2}$-$\nu f_{5/2}$
lowers the $f_{5/2}$ orbital with increasing $N$, and also provides
the mechanism of lowering the $0_{2}^{+}$ energy around $N=40$. The
attractive $\nu p_{3/2}$-$\nu g_{9/2}$ monopole interaction explains
the rapid changes of structure at $N=36$. Ge isotopes are considered
as the mirror nuclei of Cr isotopes with respect to the $Z=28$ shell
closure. We have pointed out that the systematics of energy levels
and $B(E2)$ values in Ge isotopes is quite similar to that in
neutron-rich Cr isotopes. We thus expect that our model can explain
the characteristic behaviors in Ge isotopes, and are applicable to
the $N=Z$ nuclei with nuclear mass $A=68-100$.

Y.S. is supported in part by the Chinese Major State Basic Research
Development Program through grant 2007CB815005.



\end{document}